\documentclass[prl,aps,showpacs,twocolumn,floatfix]{revtex4}
\usepackage{epsfig} \usepackage{graphics} \usepackage{bm}
\usepackage{amssymb}
\begin{document}

\title{Breakdown of the Equivalence between Energy Content and 
Weight in a Weak Gravitational Field for a Quantum Body}

\author{A.G. Lebed}

\affiliation{Department of Physics, University of Arizona, 1118 E.
4-th Street, Tucson, AZ 85721, USA}
\affiliation{L.D. Landau Institute for Theoretical Physics,
2 Kosygina Street, 117334 Moscow, Russia}
\date{\today}

\begin{abstract}
It is shown that weight operator of a composite quantum body in a weak 
external gravitational field in the post-Newtonian approximation of the 
General Relativity does not commute with its energy operator, taken in 
the absence of the field. 
Nevertheless, the weak equivalence between the expectations values of 
weight and energy is shown to survive at a macroscopic level for stationary 
quantum states for the simplest composite quantum body - a hydrogen 
atom.
Breakdown of the weak equivalence between weight and energy at a 
microscopic level for stationary quantum states can be experimentally 
detected by studying unusual electromagnetic radiation, emitted  by the 
atoms, supported and moved in the Earth gravitational field with constant
velocity, using spacecraft or satellite. 
For superpositions of stationary quantum states, a breakdown of the above 
mentioned equivalence at a macroscopic level leads to time dependent 
oscillations of the expectation values of weight, where the equivalence 
restores after averaging over time procedure.
\end{abstract}

\maketitle

\pagebreak

Formulation of a successful quantum gravitation theory is considered
to be one of the most important problems in physics and the major
step towards the so-called 
"Theory of Everything". 
On the other hand, fundamentals of the General Relativity (GR) and 
quantum mechanics are so different that there is a possibility that it 
will not be possible to unite these two theories in a feasible 
future. 
In this difficult situation, it seems to be important to suggest a
combination of the quantum mechanics and some non-trivial
approximation of the GR. 
In particular, this is important in case, where such theory can be 
experimentally tested. 
To the best of our knowledge, so far only quantum variant of trivial 
Newtonian approximation of the GR has been studied experimentally 
in the famous COW \cite{COW-1} and ILL \cite{ILL-1, Voronin} 
experiments.
As to such important and non-trivial quantum effects in the GR as the 
Hawking radiation \cite{Hawking} and the Unruh effect \cite{Unruh}, 
they are still very far from their direct and unequivocal experimental 
confirmations.

A notion of gravitational mass of a composite body is known to be 
non-trivial in the GR and related to the following 
paradoxes.
If we consider a free photon with energy $E$ and apply to it the 
so-called Tolman formula for gravitational mass \cite{Landau}, we will
 obtain $m^g=2E/c^2$ (i.e., two times bigger value than the expected 
 one) \cite{Misner}.
If a photon is confined in a box with mirrors, then we have a composite 
body at rest.
In this case, as shown in Ref. \cite{Misner}, we have to take into account
a negative contribution to $m^g$ from stress in the box walls to restore 
the equation $m^g=E/c^2$.
It is important that the later equation is restored only after averaging
over time.
A role of the classical virial theorem in establishing of the equivalence
between averaged over time gravitational mass and energy is discussed
in detail in Refs. \cite{Nordtvedt, Carlip} for different types of classical
composite bodies.
In particular, for electrostatically bound two bodies, it is shown
that gravitational field is coupled to a combination $3K+2U$,
where $K$ is kinetic energy, $U$ is the Coulomb potential
energy.
Since the classical virial theorem states that the following time average
is equal to zero, $\bigl< 2K+U \bigl>_t = 0$, then we conclude that
averaged over time gravitational mass is proportional to the total amount
of energy \cite{Nordtvedt, Carlip},
 \begin{equation}
\bigl< m^g \bigl>_t = \bigl<3K+2U\bigl>_t/c^2 =
\bigl<K+U\bigl>_t /c^2 = E/c^2.
\end{equation}

The main goal of our Letter is to study a quantum problem about weight 
of a composite body. 
As the simplest example, we consider a hydrogen atom in the Earth 
gravitational field, where we take into account only kinetic and Coulomb 
potential energies of an electron in a curved 
spacetime. 
We claim three main results in the Letter. 
The first our result is that the weak equivalence between weight in a
weak gravitational field and energy in the absence of the field may 
survive at a macroscopic level in a quantum case \cite{WEP}.
More strictly speaking, we show that the expectation value of the weight
is equal to $E/c^2$ for stationary quantum states due to the quantum 
virial theorem. 
The second our result is a breakdown of the weak equivalence between 
weight in a weak gravitational field and energy at a microscopic level 
for stationary quantum states due to the fact that the weight operator 
does not commute with energy operator, taken in the absence of 
gravitational field. 
As a result, there exist a non-zero probability that a measurement of
the weight gives value, which is different from
$E/c^2$.
We suggest to detect this weak inequivalence of weight in a weak
gravitational field and energy by measurements of electromagnetic 
radiation, emitted by a macroscopic ensemble of hydrogen atoms, 
supported and moved in the Earth gravitational field, by using spacecraft
or satellite \cite{Lebed-1}. 
The third our result is a breakdown of the weak equivalence between the
expectation values of the weight and energy at a macroscopic level for a 
superposition of stationary 
quantum states. 
As we show below, time dependent oscillations of the expectation 
values of the weight are expected to exist in this case, and, the equivalence 
is restored after averaging of these oscillations
over time.

Below, we derive the Lagrangian and Hamiltonian of a hydrogen atom in the Earth gravitational field, taking into account couplings of kinetic and potential Coulomb
energies of an electron with a weak gravitational
field.
Note that we keep only terms of the order of $1/c^2$ and disregard magnetic
force, radiation of both electromagnetic and gravitational waves as well as
all tidal and spin dependent effects.
Let us write the interval in the Earth gravitational field, using the so-called
weak field approximation \cite{Misner-2,Lebed-2}:
\begin{eqnarray}
d s^2 = -\biggl(1 + 2 \frac{\phi}{c^2} \biggl)(cdt)^2
+ &&\biggl(1 - 2 \frac{\phi}{c^2} \biggl) (dx^2 +dy^2+dz^2 ),
\nonumber\\
&&\phi = - \frac{GM}{R} ,
\end{eqnarray}
where $G$ is the gravitational constant, $c$ is the velocity of
light, $M$ is the Earth mass, $R$ is a distance from a center of the
Earth.

Then in the local proper spacetime coordinates,
\begin{eqnarray}
&&x'=\biggl(1-\frac{\phi}{c^2} \biggl) x, \ y'= \biggl(1-\frac{\phi}{c^2} \biggl) y,
\nonumber\\
&&z'=\biggl(1-\frac{\phi}{c^2} \biggl) z , \ t'= \biggl(1+\frac{\phi}{c^2} \biggl) t,
\end{eqnarray}
the classical Lagrangian and action of an electron in a hydrogen atom have the
following standard forms:
\begin{equation}
L' = -m_e c^2 + \frac{1}{2} m_e ({\bf v'})^2 + \frac{e^2}{r'} \ , \ \ \ S' = \int L' dt' ,
\end{equation}
where $m_e$ is the bare electron mass, $e$ and ${\bf v'}$ are the electron charge
and velocity, respectively; $r'$ is a distance between electron and
proton.
It is possible to show that the Lagrangian (4) can be rewritten in coordinates
$(x,y,z,t)$ as
\begin{equation}
L = -m_e c^2 +  \frac{1}{2}m_e{\bf v}^2+\frac{e^2}{r}
- m_e \phi
- \biggl( 3m_e\frac{{\bf v}^2}{2}-2\frac{e^2}{r} \biggl)
\frac{\phi}{c^2} .
\end{equation}

Let us calculate the Hamiltonian, corresponding to the Lagrangian (5), by means
of a standard procedure, $H({\bf p},{\bf r})={\bf p}{\bf v}-L({\bf v},{\bf r})$,
where ${\bf p}= \partial L({\bf v},{\bf r})/\partial {\bf v}$.
 As a result, we obtain:
\begin{equation}
H = m_e c^2 + \frac{{\bf p}^2}{2m_e}-\frac{e^2}{r} + m_e  \phi
+ \biggl( 3 \frac{{\bf p}^2}{2 m_e}
-2\frac{e^2}{r} \biggl) \frac{\phi}{c^2},
\end{equation}
where canonical momentum in a gravitational field is ${\bf p}=m_e
{\bf v}(1-3\phi/c^2)$.
>From the Hamiltonian (6), averaged over time electron weight in a weak
gravitational field, $<m^g_e \phi>_t$, can be expressed
as
\begin{eqnarray}
<m^g_e \phi>_t &&= m_e \phi + \biggl<  \frac{{\bf p}^2}{2 m_e}- \frac{e^2}{r}\biggl>_t \frac{\phi}{c^2}
+\biggl< 2 \frac{{\bf p}^2}{2 m_e}-\frac{e^2}{r}\biggl>_t \frac{\phi}{c^2}
\nonumber\\
&&= \biggl( m_e +
\frac{E}{c^2} \biggl) \phi \ ,
\end{eqnarray}
where $E= {\bf p}^2/2 m_e - e^2/r$ is an electron energy. 
Note that averaged over time third term in Eq.(7) is equal to zero due to the
classical virial theorem. 
Thus, we conclude that in classical physics averaged over time wieght of a 
composite body is equivalent to its energy, taken in the absence of gravitational
field \cite{Nordtvedt,Carlip}.

The Hamiltonian (6) can be quantized by substituting a momentum operator,
$\hat{\bf p} = - i \hbar \partial /\partial {\bf r}$, instead of canonical momentum,
${\bf p}$. 
It is convenient to write the quantized Hamiltonian in the
following form:
\begin{equation}
\hat H = \frac{\hat {\bf p}^2}{2m_e}-\frac{e^2}{r} + \hat m^g_e \phi \ ,
\end{equation}
where we omit term $m_e c^2$ and introduce weight operator of an electron 
in a weak gravitational
field,
\begin{equation}
\hat m^g_e \phi =m_e \phi + \biggl(\frac{\hat {\bf
p}^2}{2m_e}  -\frac{e^2}{r}\biggl)\frac{\phi}{c^2} 
+ \biggl(2 \frac{\hat {\bf p}^2}{2m_e}-\frac{e^2}{r} \biggl) \frac{\phi}{c^2} \ .
\end{equation}
Note that, in Eq.(9), the first term corresponds to the bare electron mass, 
$m_e$, the second term corresponds to the expected electron energy 
contribution to the weight operator, whereas the third non-trivial term is 
the virial contribution to the weight operator. 
It is important that the operator (9) does not commute with electron energy 
operator, taken in the absence of gravitational field. 
It is possible to show [14] that Eqs.(8),(9) can be obtained directly from 
the Dirac equation in a curved spacetime, corresponding to a weak
gravitational field (2).

Below, we discuss some consequences of Eqs.(8),(9). Suppose that we
have a macroscopic ensemble of hydrogen atoms with each of them
being in a ground state with energy $E_1$. 
Then, from Eq.(9), it follows that the expectation value of weight operator 
per atom is
\begin{equation}
<\hat m^g_e \phi> = m_e \phi+ \frac{ E_1}{c^2} \phi + \biggl< 2 \frac{\hat
p^2}{2m_e}-\frac{e^2}{r} \biggl> \frac{\phi}{c^2} = \biggl(m_e + \frac{E_1}{c^2} 
\biggl) \phi ,
\end{equation}
where the third term in Eq.(10) is zero in accordance with the
quantum virial theorem \cite{Virial}. 
Therefore, we conclude that the weak equivalence between weight in a 
weak gravitational field and energy in the absence of the field survives at 
a macroscopic level for stationary quantum 
states.

Let us discuss how Eqs.(8),(9) break the weak equivalence between
weight in a weak gravitational field and energy at a 
microscopic level. 
First of all, we pay attention that the weight operator (9) does not commute 
with electron energy operator, taken in the absence of 
gravitational field. 
This means that, if we create a quantum state of a hydrogen atom with 
definite energy, it will not be characterized by definite weight. 
In other words, a measurement of the weight in such quantum state may 
give different values, which, as shown, are
quantized. 
Here, we illustrate the above mentioned inequivalence,
using the following thought experiment. Suppose that at $t=0$ we
create a ground state wave function of a hydrogen atom,
corresponding to the absence of gravitational field,
\begin{equation}
\Psi_1(r,t) = \Psi_1(r) \exp(-iE_1t/\hbar) \ .
\end{equation}
In a weak gravitational field (2), wave function (11) is not anymore a ground
state of the Hamiltonian (8),(9) from point of view of an inertial observer,
located at infinity.
For such observer, in accordance with Eq.(3), a general solution of the Schrodinger
equation, corresponding to the Hamiltonian (8),(9), can be written as
\begin{equation}
\Psi(r,t)
= \sum^{\infty}_{n = 1} a_n \Psi_n [(1- \phi/c^2)r] \exp[-i E_n(1+\phi/c^2) t/\hbar] \ .
\end{equation}
[Here factor $1-\phi/c^2$ is due to a curvature of space, whereas the term
$E_n(1+\phi/c^2)$ reflects the famous red shift in gravitational field and is due
to a curvature of time.
$\Psi_n(r)$ is a normalized wave function of an electron in a hydrogen
atom in the absence of gravitational field, corresponding
to energy $E_n$ [16].]

In accordance with the quantum mechanics, probability that at $t>0$ an electron
occupies excited state with energy $E_n(1+\phi/c^2)$ is
\begin{eqnarray}
&&P_n = |a_n|^2, \ a_n = \int \Psi^*_1(r) \Psi_n [(1-\phi/c^2)r] d^3 {\bf r}
\nonumber\\
&&= - ( \phi/c^2) \int \Psi^*_1(r) r \Psi'_n(r) d^3 {\bf r}, \  \ n \neq 1 .
\end{eqnarray}
Taking into account that the Hamiltonian is the Hermitian operator, it is possible to
show that
\begin{equation}
\int \Psi^*_1(r) r \Psi'_n(r) d^3 {\bf r} = \frac{V_{n,1}}{\hbar \omega_{n,1}}, \ \
\hbar \omega_{n,1} = E_n-E_1 ,
\end{equation}
where
\begin{equation}
V_{n,1}= \int \Psi^*_1(r) \hat V({\bf r}) \Psi_n(r) d^3 {\bf r} , \ \
\hat V({\bf r}) = 2 \frac{\hat {\bf p}^2}{2 m_e} - \frac{e^2}{r}.
\end{equation}

Let us discuss Eqs.(12)-(15).
Note that they directly demonstrate that there is a finite probability,
\begin{equation}
P_n = |a_n|^2 = \Big( \frac{\phi}{c^2} \Big)^2 \
\Big(  \frac{V_{n,1}}{E_n-E_1} \Big)^2 \ , \ n \neq 1,
\end{equation}
that at $t>0$ an electron occupies n-th energy level. 
In fact, this means that measurement of weight in a weak gravitational field
in a quantum state with a definite energy (11) gives the following
quantized values:
\begin{equation}
m^g_e (n) \phi = m_e \phi + (E_n/c^2) \phi \ ,
\end{equation}
corresponding to the probabilities (16) [17].
[Note that $\hat V({\bf r})$ in Eq.(15) is the virial operator. 
It is a part of the weight operator (9), which does not commute with energy 
operator, taken in the absence of gravitational field. 
Due to the fact that $\hat V({\bf r})$ presents in Eqs.(9),(15), the probabilities 
(16) for the quantization law (17) are not equal to zero.]
We point out that, although the probabilities (16) are quadratic with respect
to gravitational potential and, thus, small, the changes of the weight  (17) 
are large and of the order of $\alpha^2 m_e$, where $\alpha$ is the fine 
structure constant.
We also pay attention that small values of probabilities (16), $P_n \sim 10^{-18}$,
do not contradict to the existing Eotvos type measurements [12], which have
confirmed the weak equivalence principle with the accuracy of the order of 
$10^{-12}-10^{-13}$. 
For us, it is very important that the excited levels of a hydrogen atom spontaneously 
decay with time, therefore, one can detect quantization law (17) by measuring
electromagnetic radiation, emitted by a macroscopic ensemble 
of hydrogen atoms.
The above mentioned optical method is much more sensitive than the Eotvos 
type measurements and we, therefore, hope that it will allow to detect the breakdown
of  the equivalence between energy content and weight in a weak gravitational field, 
suggested in the Letter. [For more details, see the description of a realistic experiment 
below.]

Here, we describe a realistic experiment \cite{Lebed-1}. We consider
a hydrogen atom to be in its ground state at $t=0$ and located at
distance $R'$ from a center of the Earth. 
The corresponding wave function can be written as
\begin{equation}
\tilde{\Psi}_1(r,t) = \Psi_1[(1-\phi'/c^2)r] \exp[-iE_1(1+\phi'/c^2)t/\hbar] \ ,
\end{equation}
where $\phi'=\phi(R')$.
The atom is supported in the Earth gravitational field and moved from the Earth
with constants velocity, $v \ll \alpha c$, by spacecraft 
or satellite.
As follows from Ref.[8], the extra contributions to the Lagrangian (5) are small in this case in an inertial system, related to a hydrogen
atom.
Therefore, electron wave function and time dependent perturbation for the
Hamiltonian (8),(9) in this inertial coordinate system can be expressed as [18]
\begin{equation}
\tilde{\Psi}(r,t) = \sum^{\infty}_{n=1} \tilde{a}_n(t) \Psi_n[(1-\phi'/c^2)r] 
\exp[-iE_n(1+\phi'/c^2)t/\hbar] ,
\end{equation}
\begin{equation}
\hat U ({\bf r},t) =\frac{\phi(R'+vt)-\phi(R')}{c^2}  \biggl(3 \frac{\hat {\bf p}^2}{2m_e}-2\frac{e^2}{r} \biggl) .
\end{equation}
Application of the time-dependent quantum mechanical perturbation theory
gives the following solutions for functions $\tilde a_n(t)$ in Eq.(19):
\begin{equation}
\tilde{a}_n(t)= \frac{\phi(R')-\phi(R'+vt)}{c^2} \frac{V_{n,1}}{\hbar \omega_{n,1}}
\exp(i \omega_{n,1})\  , \ n \neq 1 \ ,
\end{equation}
where $V_{n,1}$ and $\omega_{n,1}$ are given by Eqs.(14),(15);
$\omega_{n,1} \gg v/R'$.

It is important that, if excited levels of a hydrogen atom were strictly stationary, 
then a probability to find the weight to be quantized with $n \neq 1$ (17) 
would be
\begin{equation}
\tilde{P}_n(t)= \biggl( \frac{V_{n,1}}{\hbar \omega_{n,1}} \biggl)^2
 \frac{[\phi(R'+vt)-\phi(R')]^2}{c^4}   \ , n \neq 1.
\end{equation}
In reality, the excited levels spontaneously decay with time and, therefore, it is possible to observe the quantization law (17) indirectly by measuring
electromagnetic radiation from a macroscopic ensemble of
the atoms.
In this case, Eq.(22) gives a probability that a hydrogen atom emits a photon
with frequency $\omega_{n,1} = (E_n-E_1) / \hbar$ during the time interval
$t$ [19].

Let us estimate the probability (22). If the experiment is done by
using spacecraft or satellite, then we may have $|\phi(R'+vt)|  \ll
|\phi(R')|$. In this case Eq.(22) is reduced to Eq.(16) and can be
rewritten as
 \begin{equation}
\tilde{P}_n = \biggl( \frac{V_{n,1}}{E_n - E_1} \biggl)^2
 \frac{\phi^2(R')}{c^4}  \simeq  0.49 \times 10^{-18} \biggl( \frac{V_{n,1}}{E_n-E_1} \biggl)^2 ,
\end{equation}
where, in Eq.(23), we use the following  numerical values of the
Earth mass, $M \simeq 6 \times 10^{24} kg$, and its radius, $R_0
\simeq 6.36 \times 10^6 m$. Note that although the probabilities (23)
are small, the number of photons, $N$, emitted by macroscopic
ensemble of the atoms can be large since the factor
$V^2_{n,1}/(E_n-E_1)^2$ is of the order of unity. For instance, for
1000 moles of hydrogen atoms, $N$ is estimated as
\begin{eqnarray}
&&N (n \rightarrow 1) = 2.95 \times 10^{8} \biggl( \frac{V_{n,1}}{E_n-E_1} \biggl)^2 ,
\nonumber\\
&&N (2 \rightarrow 1) = 0.9 \times 10^8 ,
\end{eqnarray}
which can be hopefully experimentally detected.
[Here $N(n \rightarrow 1)$ stands for a number of photons, emitted with energy
$\hbar \omega_{n,1} = E_n -E_1$.]

To summarize, we have demonstrated that weight of a composite quantum 
body in a weak external gravitational field is not equivalent to its energy in 
the weak sense due to quantum fluctuations and discussed a possible
indirect experimental method to detect this difference. 
We have also shown that the corresponding expectation values are 
equivalent to each other for stationary quantum states. 
In this context, we need to make the following comment. 
First of all, we stress that, for superpositions of stationary states, the 
expectation values of the weight  can be oscillatory functions of time 
even in case, where the expectation value of energy is 
constant. 
For instance, as follows from Eq.(9), for electron wave function,
\begin{equation}
\Psi_{1,2}(r,t) = \frac{1}{\sqrt{2}} \bigl[ \Psi_1(r) \exp(-iE_1t) + \Psi_2(r) 
\exp(-iE_2t) \bigl],
\end{equation}
which is characterized by the time independent expectation value of
energy, $<E> = (E_1+E_2)/2$, the expectation value of electron weight
is the following oscillatory function [20]:
\begin{equation}
<\hat m^g_e \phi> = m_e \phi + \frac{E_1+E_2}{2 c^2} \phi + \frac{V_{1,2}}{c^2}
\phi \cos \biggl[ \frac{(E_1-E_2)t}{\hbar} \biggl] .
\end{equation}
Note that the oscillations of the weight (26) directly demonstrate 
inequivalence of the weight and energy at a macroscopic 
level. 
It is important that these oscillations are strong (of the order of $\alpha^2 m_e$) 
and of a pure quantum origin without classical analogs. 
We hope that the above mentioned oscillations of the weight are experimentally measured, despite the fact that the quantum state (25) decays 
with time.

If we average the oscillations (26) over time, we obtain the
modified weak equivalence principle between the averaged over time
expectation value of the weight and the expectation value of energy in the 
following form:
\begin{equation}
<< \hat m^g_e \phi >>_t  = m_e \phi + \frac{(E_1+E_2)}{2c^2} \phi .
\end{equation}
We pay attention that physical meaning of averaging procedure in Eq.(27)
is completely different from that in classical time averaging procedure (1)
and does not have the corresponding classical analog.

In conclusion, we stress that we have considered in the Letter a point-like 
[21] composite quantum test body and all our results are due to 
different couplings of kinetic and potential energies with an external gravitational
field. 
This physical mechanism is completely different from those, considered before 
[22-26], where a possibility of a breakdown of the weak equivalence principle was discussed due to three mass dependent phenomena: penetration of the de Broglie waves in classically restricted areas, bound states of particles in an external
gravitational field, and the interference of the de Broglie waves.
In addition, we point out that there exists an alternative point of view (see, for 
example, Refs.[23,27]), stating that there cannot be violations due to quantum 
effects of some generalized weak equivalence principle in any metric theory of gravitation, including
the GR.

We are thankful to N.N. Bagmet for useful discussions. This work was
supported by the NSF under Grants DMR-0705986 and DMR-1104512.


\begin{references}
\bibitem{WEP}
As usual, in a framework of the weak equivalence principle, 
we do not take into account gravitational field of a test body 
(i.e., a hydrogen atom).
\bibitem{Lebed-1}
In practice, it may be more convenient to investigate electromagnetic
radiation from some solid body.
\bibitem{Lebed-2}
We pay attention that to calculate the Lagrangian (and later - the Hamiltonian)
in a linear with respect to a small parameter $\phi(R) / c^2$ approximation,
we do not need to keep the terms of the order of $[\phi(R)/c^2]^2$ in metric
(2), in contrast to the perihelion orbit procession calculations \cite{Misner-2}.
\bibitem{Lebed-4}
Note that, due to symmetry of our problem, an electron from $1S$ ground
state of a hydrogen atom can be excited only into $nS$ excited states.
We also pay attention that the wave function (12) contains a normalization 
factor $(1-\phi/c^2)^{3/2}$. 
We omit this factor in Eq.(12) and everywhere below, since it does not change
our results in a weak field approximation, corresponding to the interval (2).
\bibitem{Lebed-52}
Note that, as it follows from quantum mechanics, we have to 
calculate wave function (12) in a linear approximation with respect to
the parameter $\phi/c^2$ to obtain probabilities (16),(22),(23), which are
proportional to $(\phi/c^2)^2$.
A simple analysis shows that an inclusion in Eq.(12) terms of the order of
$(\phi/c^2)^2$ would change electron weight of the order of  $(\phi / c^2) 
m_e \phi \sim 10^{-9} m_e \phi$, which is much smaller than the distance
between the quantized values (17), $\delta m^g_e \phi \sim \alpha^2 m_e \phi
\sim 10^{-4} m_e \phi$.
\bibitem{Lebed-5}
We pay attention that dipole matrix elements for $nS \rightarrow 1S$
quantum transitions are zero. Nevertheless, the corresponding photons can be emitted due to quadrupole effects.
\bibitem{Lebed20}
We use the term point-like body in a sense that the Bohr radius,
characterizing a typical size of electron wave functions in
a hydrogen atom, is much less than both $vt$ and $R_0$. Therefore, we
disregard all small tidal effects.
\end{references}
\end{document}